          [1999/12/02 1.01 (PWD)]
\documentclass[letterpaper,twocolumn]{esapub} % American paper

\usepackage{natbib}
\usepackage{graphicx}

\title{Simulated Performance of a Germanium Compton Telescope}
\author[1,2]{Steven E. Boggs}
\affil[1]{Space Radiation Laboratory, California Institute of Technology,
MC 220-47, Pasadena, CA 91125 USA}
\affil[2]{Department of Physics, University of California,
Berkeley, CA 94720 USA}
\author[3]{Pierre Jean}
\affil[3]{Centre d'Etude Spatiale des Rayonnements, UPS-CNRS, Toulouse, France}

\begin{document}


\maketitle

\begin{abstract}
To build upon the goals of the upcoming INTEGRAL mission, the next
generation soft $\gamma$-ray (0.2-20~MeV) observatory will require
improved sensitivity to nuclear line emission while maintaining
high spectral resolution. We present the simulated performance of
a germanium Compton telescope (GCT) design, which will allow a factor
of ten improvement in sensitivity over INTEGRAL/SPI. We also
discuss a number of issues concerning reconstruction techniques
and event cuts, and demonstrate how these affect the overall
performance of the telescope.
\end{abstract}

\section{Introduction}
Combining the success of
COMPTEL/CGRO and the spectral resolution of INTEGRAL/SPI, a number
of researchers \citep{johns96, jean96, boggs98} have discussed
the merits of a germanium Compton telescope (GCT).
Compton telescopes work on a well-known principle: by measuring
the positions and energies of the photon interactions the
initial photon direction can be reconstructed to within an annulus
on the sky using the Compton scatter formula (Figure 1).
The uncertainty, or width, of this annulus depends on the spatial
and spectral resolution of the detectors, but also has a fundamental
limit set by Doppler broadening due to Compton scattering off
of bound electrons. Reconstruction of the event annulus requires that
the first and second photon interaction locations in the instrument are
spatially resolved, and their order properly determined. 

Germanium detectors pose two major complications
for Compton telescope designs. Photons above $\sim$0.5~MeV predominantly
scatter multiple times in germanium before being photoabsorbed. Also,
the expected event timing resolution in germanium detectors ($>$10~ns)
is not adequate to determine the interaction order for reasonable
GCT configurations.

\begin{figure}
\centering
\includegraphics[width=1.0\linewidth]{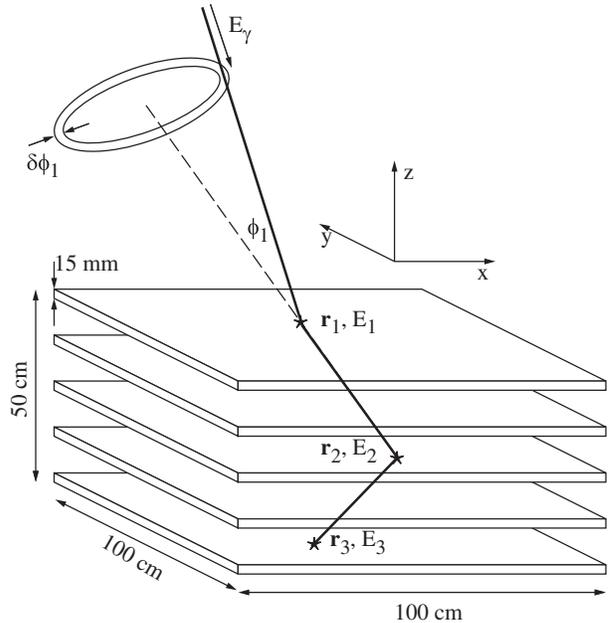}
\caption{Germanium Compton telescope configuration
analyzed in this work.\label{figure1}}
\end{figure}

In a previous paper \citep{boggs00}, hereafter Paper I, we introduced
two \textit{reconstruction techniques} to accurately
determine the photon interaction order in GCTs.
The first technique, Compton Kinematic Discrimination (CKD), takes advantage
of redundant information for photons which interact three or more times
in the instrument (3+ site events) to determine the most probable
interaction order. CKD additionally allows efficient
rejection of background events, including photons which scatter out of the
instrument before fully depositing their energy (Compton
continuum photons), non-localized $\beta^{-}$-decays, $\beta^{+}$-decays, and
pair-production events. The second reconstruction technique,
Single Scatter Discrimination (SSD), allows good determination of
the interaction order in 2-site events, but without the
benefit of background rejection.
(Also in Paper I we summarized additional techniques
for rejecting $\beta^{-}$-decays, $\beta^{+}$-decays, and
pair-production events.)
In addition to CKD and SSD, other reconstruction
techniques can be imagined.
As we discussed in Paper I, for a given GCT configuration
the performance will depend on the reconstruction techniques employed.

Telescope performance will also depend on
\textit{event cuts}, which can be made on the initial direction of the
photon scatter, the number of interaction
sites, and the minimum separation between
the first and second interaction sites (\textit{minimum lever arm}).
The tradeoffs are generally higher efficiency at the expense of
degraded angular resolution, and hence increased background.
First, the uncertainty in the Compton scatter angle (angular resolution)
is smaller for forward scatter events than backscatter events (Paper I,
Equation 4). Second, events with only
2 interaction sites (2-site) do not permit CKD background rejection,
and also have a larger fraction of backscatter events that 3+ site
events. Finally, a larger minimum lever arm will minimize the effects of
spatial uncertainty in the detectors, improving angular resolution
and hence background, but at the expense of lower efficiency.

%\begin{figure}
%\centering
%\includegraphics[width=0.9\linewidth]{figure2.eps}
%\caption{The on-axis ARM distribution at 1 MeV for Cases 1-4
%presented in Table 1 and Section 3.\label{figure2}}
%\end{figure}

\begin{figure}
\centering
\includegraphics[width=0.9\linewidth]{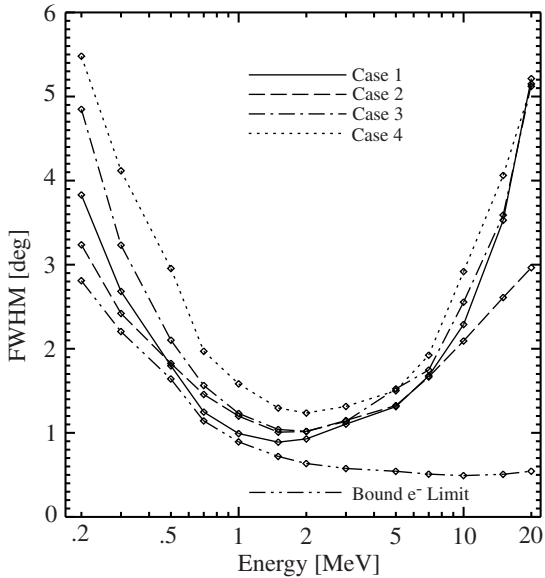}
\caption{Angular resolution for on-axis sources as
a function of photon energy. Also shown is the limit set
by Doppler broadening of bound electrons.\label{figure3}}
\end{figure}

Here we present detailed simulations of a
GCT configuration in an effort to determine the optimized sensitivity to
nuclear line emission, as well as demonstrate the
variation in performance for several different combinations of
reconstruction techniques and event cuts.
Our selections range from utilizing most of the event
information to physically reconstruct the event (Case 1, with CKD), to using
a purely empirical, but highly efficient approach (Case 4), and
should fairly represent the range of performance
characteristics possible within a GCT.

\begin{figure}
\centering
\includegraphics[width=0.9\linewidth]{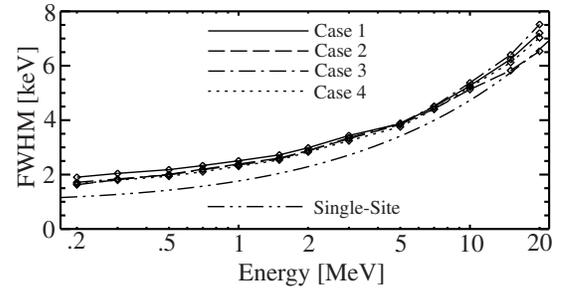}
\caption{Photopeak spectral resolutions, which show
little dependence on the reconstruction technique or event cuts.
For comparison is shown the assumed single-site resolution.\label{figure4}}
\end{figure}

\section{Telescope Simulation}

The telescope 
configuration modeled in this study is presented in Figure~1. The instrument 
consists of five planar arrays of 15~mm thick germanium, each of area 
$100~cm \times 100~cm$. In reality each array would consist of
separate smaller 
detectors $(\sim 5~cm \times 5~cm)$ tiled to form the entire plane;
however, the 
simulation performed here modeled each plane as a solid detector for 
simplicity. The five planar arrays are vertically spaced 12.5~cm
apart center-to-center.

The instrument was simulated using CERN's GEANT Monte Carlo code,
modified to include Doppler broadening of
bound electrons\footnote{The authors are grateful to R. M. Kippen for
providing his modifications to the GEANT3 software
package, which include the effects of scattering off of bound electrons.
These modifications are available on-line.
\texttt{http://gammaray.msfc.nasa.gov/actsim/}}.
Photon interactions are randomly modified to
reflect the assumed energy resolution (shown in Figure 4),
and $\sim$1-mm spatial resolution.
(These uncertainties were described in detail in Appendix A of Paper I.)
Unresolved interactions are combined, and a
detector threshold of 10~keV is assumed.

Components inducing soft $\gamma$-ray background in spaceborne instruments
are the cosmic diffuse background
(CDB) and the cosmic-ray protons that either promptly release
their energy in the detector (CR prompt), or create radioactive nuclei
in the instrument materials (CR delayed). The spacecraft is also a
source of background events since under CR or CDB irradiation, it
generates secondary particles (p$^+$, n, $\gamma$) able to reach the
instrument. Therefore, a numerical model of a spacecraft has been added
underneath the telescope presented in figure 1 in order to estimate a
more realistic background. We simulate the
irradiation of the GCT and the spacecraft by cosmic-ray fluxes in HEO
condition using the GEANT/GCALOR code. The CDB input spectrum
is based on the recent
measurements by COMPTEL \citep{weiden99} and SMM \citep{watan97}.
The solar maximum CR spectrum \citep{webber74} has been used to simulate
the CR components. The yields of radioactive nuclei induced by CRs (and
their secondaries) in the Ge have been used to calculate the decay
rates after one year in orbit. Using these rates and the ENSDF
database, the radioactive decays have been simulated with GEANT.

\section{Reconstruction Techniques \& Event Cuts}

While CKD offers a powerful technique for background rejection,
it also rejects a large fraction of signal photons which fully deposit
their energy in the instrument but, for example, have two or more
interaction sites not spatially resolved.

\begin{figure}
\centering
\includegraphics[width=0.9\linewidth]{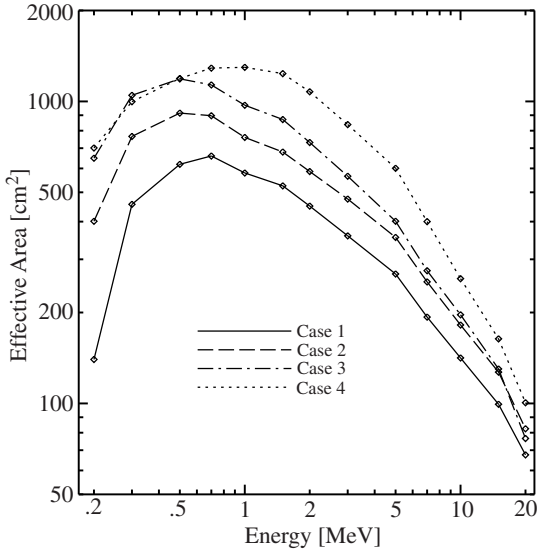}
\caption{On-axis photopeak effective areas, which show
a strong dependence on the reconstruction
technique and event cuts.\label{figure5}}
\end{figure}

As an alternative to CKD and SSD, more empirical
reconstruction techniques to determine the interaction order
can be imagined.
For example, the Monte Carlo simulations show that
for 1 MeV photons, the majority of events have their largest
energy deposit in the first (70.2\%) or second (22.7\%)
interaction site. This information allows us to devise a different
reconstruction technique. First we define a
source position on the sky. Then we assume that the largest energy deposit for
a given event is either the first or second interaction site, for a photon
of energy equal to the total energy deposited in the instrument.
Then we test each other interaction site in combination with this largest
energy deposit to determine if any pair is
consistent with the first and second interactions of a photon originating
from the defined source position. Events which
are not consistent with this source position are rejected.
This empirical reconstruction technique
dramatically increases the effective area relative to CKD, but also
increases the background and Compton continuum components.

\begin{figure}
\centering
\includegraphics[width=0.9\linewidth]{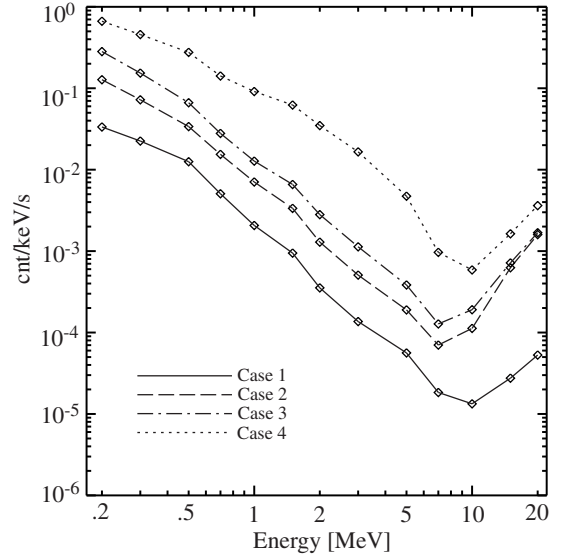}
\caption{Background events consistent with the 78\% error circle of
an on-axis point source.
\label{figure6}}
\end{figure}

The four (4) combinations of reconstruction techniques and event cuts
presented here are summarized in Table 1.
These cases are roughly ordered from highest angular resolution and
lowest background (Case 1), to highest effective area (Case 4).
In addition to CKD in Cases 1-3, we have included in all four cases the additional
background rejection techniques discussed in Paper I
(e.g., rejection of events with $\beta^{-}$-signatures,
positron signatures, 8+ site events).
Case 4 uses the empirical reconstruction technique discussed above.
The 10-cm minimum lever arm in Cases 1 \& 2
almost always requires the first and second interactions to be
in separate detector planes, while the 1-cm in Cases 3 \& 4 allows
these interactions in the same plane to increase efficiency, but at the
cost of degraded angular resolution and higher background.

\begin{table}
  \begin{center}
    \caption{Reconstruction techniques and event cuts presented.}
    \begin{tabular}[h]{lcccc}
      \hline
      Case & Ordering  & $\#$  & Back-    & Minimum \\
           & Technique & Sites & scatters & Lever Arm \\
      \hline
      1  & CKD & 3+ & no & 10~cm \\
      2  & CKD/SSD & 2+ & yes & 10~cm \\
      3  & CKD/SSD & 2+ & yes & 1~cm \\
      4  & Empirical & 2+ & yes & 1~cm \\
      \hline \\
      \end{tabular}
    \label{table1}
  \end{center}
\end{table}

\section{Performances}

Angular resolution in Compton telescopes is often described in terms of the
\textit{angular resolution measure} (ARM),
defined as the difference between the initial
photon scatter angle in the instrument and the scatter angle reconstructed from
the Compton scatter formula.
%The ARM distribution at 1.0~MeV
%for on-axis sources is shown in Figure 2.
%The progressive broadening of the ARM distribution
%from Case 1 to Case 4 is clear.
The ARM distributions are highly non-Gaussian,
with sharp central peaks and broad wings. The FWHM of the on-axis
ARM distribution as a function of energy is shown in Figure 2, as
well as the limits imposed by Doppler broadening of bound electrons,
which dominates the angular resolution below $\sim$2~MeV.
At higher energies, the angular resolution is dominated by the degradation
in the spatial resolution due to the increased recoil-electron range.
The overall degradation at 1~MeV is 58\% between Cases 1 and 4
(0.96$^{\circ}$ to 1.52$^{\circ}$).

%\begin{table}
%  \begin{center}
%    \caption{Performance at 1~MeV.}
%%    \renewcommand{\arraystretch}{1.2}
%    \begin{tabular}[h]{ccccc}
%      \hline
%      Case & 1 & 2  & 3 & 4 \\
%      \hline
%      Effective Area $[cm^{2}]$ & 590 & 770 & 980 & 1330 \\
%      ARM FWHM [deg] & 0.96 & 1.16 & 1.17 & 1.52 \\
%      $\Delta$E FWHM [keV] & 2.50 & 2.36 & 2.35 & 2.26 \\
%      Narrow Line Sensitivity & 5.4 & 7.4 & 7.9 & 16.3 \\
%      $[10^{-7} ph/cm^{2}/s]$ &  &  &  &  \\
%      \hline \\
%      \end{tabular}
%    \label{table2}
%  \end{center}
%\end{table}

The photopeak FWHM energy resolution is broader than
the single-site resolution due to the addition
in quadrature of electronic noise for multiple interaction sites.
The photopeak energy resolution is shown in Figure~3, with the assumed single-site
resolution shown for comparison. The photopeak resolution is nearly identical
in Cases 1-4. At 1 MeV, the photopeak FWHM corresponds to a
$\sim$33\% broadening over the single-site resolution (1.77~keV to 2.36~keV).

The on-axis photopeak effective area as a function of energy is shown in Figure~4,
determined by integrating over the ARM distribution.
The effective area peaks around 0.5-1.0~MeV, with considerable area between
0.2 and 20~MeV, and is sensitive to the reconstruction technique and event cuts.
Between Cases 1 and 4, the overall increase in effective area at 1 MeV is 125\%
(590~cm$^{2}$ to 1330~cm$^{2}$).

\begin{figure}
\centering
\includegraphics[width=0.9\linewidth]{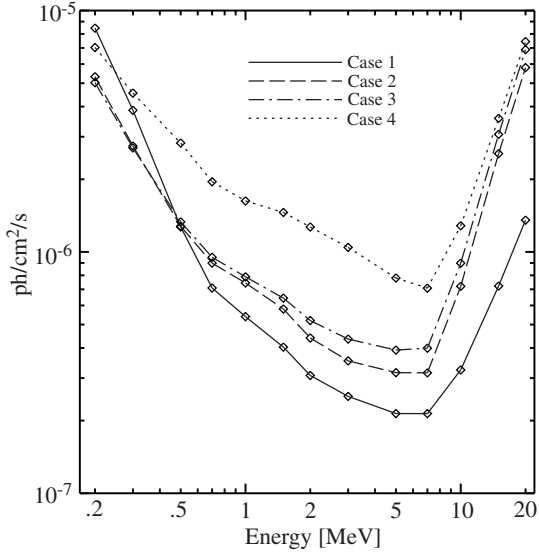}
\caption{The narrow-line sensitivity (3$\sigma$,10$^{6}$sec), for
an on-axis point source.
\label{figure7}}
\end{figure}

The total background for an on-axis point source observation is
shown in Figure 5.
This background is defined as all events whose event circles are
consistent with a photon originating from within the 78\% error
circle (\textit{effective FWHM}) of the energy-dependent ARM
distribution for each particular case. At 1~MeV, the background
is a factor of 40 higher in Case 4 than in Case 1. This large
variation is due to the increased effective area and degraded angular
resolution, as well as the minimal background rejection, in Case 4.
The corresponding  narrow-line sensitivities (3$\sigma$,10$^{6}$sec)
are shown in Figure~6. At 1~MeV, Case 1 has the highest sensitivity 
at $5.1 \times 10^{-7}$ ph/cm$^{2}$/s, degrading to
$15.0 \times 10^{-7}$ ph/cm$^{2}$/s in Case 4. We have not accounted
for instrument deadtimes when processing photon or charged-particle
events; however, we estimate that deadtime will affect the sensitivity
by $\ll 5\%$.

\section{Discussion}

GCTs offer an attractive option for a soft $\gamma$-ray
observatory
following INTEGRAL. The GCT presented here will allow a factor of
10 improvement in nuclear line sensitivity over INTEGRAL,
which is required for new scientific goals such as the systematic
study of Type Ia SNe, as well as improved imaging of
the positron annihilation line, $^{26}$Al, $^{60}$Fe, and $^{44}$Ti.
We have also demonstrated how a judicious choice of
reconstruction techniques (CKD) and event cuts
(3+ site, forward scatters) results in a 3-fold improvement in sensitivity
near 1~MeV, while maintaining 
angular resolution near the Doppler-broadening limit.
Our next goal is to study the performance of several different
GCTs configurations, varying the detector-plane spacings, to determine how
the geometry affects performance.

\section*{Acknowledgments}

S. Boggs acknowledges the Caltech Millikan Fellowship Program
and NASA Grant NAT5-5285 for support. Thanks to R.~M.~Kippen for
providing useful information on bound-electron Doppler broadening.

% The following bibliography was produced with
%   \bibliographystyle{aa}
%   \bibliography{esapub}
% The results are inserted directly here to simplify
% the demonstration.

\end{document}